\documentclass[twocolumn,prl,floatfix,citeautoscript,nofootinbib,superscriptaddress]{revtex4-2}
\usepackage{amsbsy}
\usepackage{latexsym,epsfig,graphicx}
\usepackage{dcolumn}
\usepackage{graphicx}
\usepackage{subfigure}
\usepackage{comment}
\usepackage{color}
\usepackage{bm}
\usepackage{mathrsfs}
\usepackage{amsfonts}
\usepackage{amsmath}
\usepackage{color}
\usepackage{amssymb}
\usepackage{xspace}
\usepackage{epstopdf}
\usepackage{tabularx}
\usepackage{longtable}
\usepackage[colorlinks=true, letterpaper=true, pdfstartview=FitV, linkcolor=blue, citecolor=blue, urlcolor=blue]{hyperref}
\usepackage[normalem]{ulem}
\usepackage{multirow}
\usepackage{esint}
\usepackage{mathdots}
\usepackage{soul}

\allowdisplaybreaks[4]
\pdfoutput=1

\begin{document}
\title{Virus Spreading in Quantum Networks}

\author{Junpeng Hou}
\email{jhou@pinterest.com}
\thanks{These authors contributed equally to this work.}
\affiliation{Pinterest Inc., San Francisco, California 94103, USA}

\author{Mark Maximilian Seidel}
\thanks{These authors contributed equally to this work.}
\affiliation{Department of Electrical and Systems Engineering, Washington University in St. Louis, St. Louis, MO 63130, USA}

\author{Chuanwei Zhang}
\email{chuanwei.zhang@wustl.edu}
\affiliation{Department of Physics, Washington University in St. Louis, St. Louis, MO 63130, USA}
\affiliation{Center for Quantum Leap, Washington University in St. Louis, St. Louis, MO 63130, USA}

\begin{abstract}
Recent advances in quantum communication have enabled long-distance secure information transfer through quantum channels, giving rise to quantum networks with unique physical and statistical properties. However, as in classical networks, the propagation of viruses in these systems could have severe consequences. Here, we investigate the critical problem of virus spreading in quantum networks. We develop quantitative tools, particularly a modified nonlinear dynamical system model, for performing epidemiological analyses on quantum networks. Our results show that quantum networks tend to be more resilient to viral infections, exhibiting higher epidemic thresholds than classical networks with identical graph topologies. This apparent robustness, however, arises primarily from the sparser connectivity inherent to the quantum networks. When the comparison is made at a fixed average connectivity, classical and quantum networks display comparable epidemic thresholds. These findings provide key insights into the security and reliability of future large-scale quantum communication systems. Our work bridges the fields of quantum information science, network theory, and epidemiology, paving the way for future studies of quantum epidemiological dynamics.
  
\end{abstract}

\maketitle

{\color{blue}\emph{Introduction.}}
The rapid advancement of quantum communication technologies is laying the groundwork for next-generation information networks \cite{GisinQuantum2007, PirandolaFundamental2017, CozzolinoHigh2019, QIRev1,QIRev2, SidhuAdvances2021, Chenintegrated2021, ValivarthiQuantum2016, SunQuantum2016, SimonTowards2017, LiaoSatellite2018}. Quantum network, an interconnected network capable of transmitting quantum information over long distances, holds transformative potential for secure communication, distributed quantum computing, and large-scale entanglement distribution \cite{QIRev1,QIRev2}. The road to quantum networks builds on key milestones such as quantum key distribution (QKD), which provides information-theoretic security \cite{ShorSimple2000, BarrettSignaling2005, Scaranisecurity2009}, and quantum teleportation, which enables the transfer of quantum states between distant nodes \cite{QIRev1,GisinQuantum2007,BouwmeesterExperimental1997}. Over the past decade, experimental breakthroughs in satellite-based QKD and long-distance optical links have brought the vision of a large-scale quantum network closer to reality \cite{LiaoSatellite2017, XuSecure2020, Zhangdevice2022, NadlingerExperimental2022, RiebeDeterministic2004, YonezawaDemonstration2004, ShersonQuantum2006, MaQuantum2012, PirandolaAdvances2015, RenGround2017}. Alongside hardware progress, extensive research on network architectures \cite{MehicQuantum2020}, quantum networks protocols \cite{KozlowskiDesigning2020, BritoQuantum2022},  distributed quantum computing \cite{CacciapuotiQuantum2020}
, and statistical properties \cite{BritoStatistical2020} has deepened our understanding of how quantum effects reshape the structure and functionality of complex networks.

As with classical communication networks, quantum networks may also face risks of viral propagation that threaten both the security and stability of communication systems, potentially leading to widespread disruptions. A quantum virus can be viewed as the quantum counterpart of a computer virus, i.e., an unwanted process that spreads through a quantum network and interferes with communication or computation. Research on quantum viruses in the context of quantum computation and communication is still in its infancy. At present, there is no well-established definition or systematic understanding of their mechanisms, behaviors, or potential consequences. For instance, a quantum virus could arise from a malicious circuit that interacts with a computational circuit through cross-talk \cite{Deshpande2023}, or from compromised quantum photons used in QKD, where infections at the classical control layers introduce misleading information into the system. Developing a clear understanding of how such viruses propagate in quantum networks is therefore crucial for safeguarding the security and reliability of future quantum communication infrastructures.

Despite extensive studies of virus dynamics in classical networks \cite{BorgattiNetwork2011}, ranging from the Kephart–White model \cite{KephartDirected1991} to mean-field \cite{SatorrasEpidemic2002} and nonlinear dynamical frameworks \cite{WangEpidemic2003, MieghemVirus2009, MieghemInhomogeneous2013}, little attention has been given to how uniquely quantum features influence epidemic behavior. The lack of statistical analyses that account for quantum connectivity, entanglement correlations, and probabilistic communication pathways leaves a significant gap in our understanding of network vulnerabilities in the quantum regime.

In this work, we take the first step toward studying virus propagation in quantum networks, focusing on the statistical characteristics of viral dynamics in QKD systems. We extend several classical epidemic models to incorporate the structural and statistical features of quantum network architectures and develop a modified nonlinear dynamical system (mNLDS) framework to describe their infection dynamics. Our analysis shows that quantum networks exhibit higher epidemic thresholds than classical counterparts with identical topologies, indicating enhanced resilience. However, this apparent robustness primarily arises from the sparser connectivity typical of quantum networks. When the comparison is made at equal average connectivity, the difference between classical and quantum epidemic thresholds becomes negligible. These results establish a foundational framework for quantum epidemiology and provide valuable insights for designing secure and reliable large-scale quantum communication networks.

\begin{figure}
	\includegraphics[width=0.97\columnwidth]{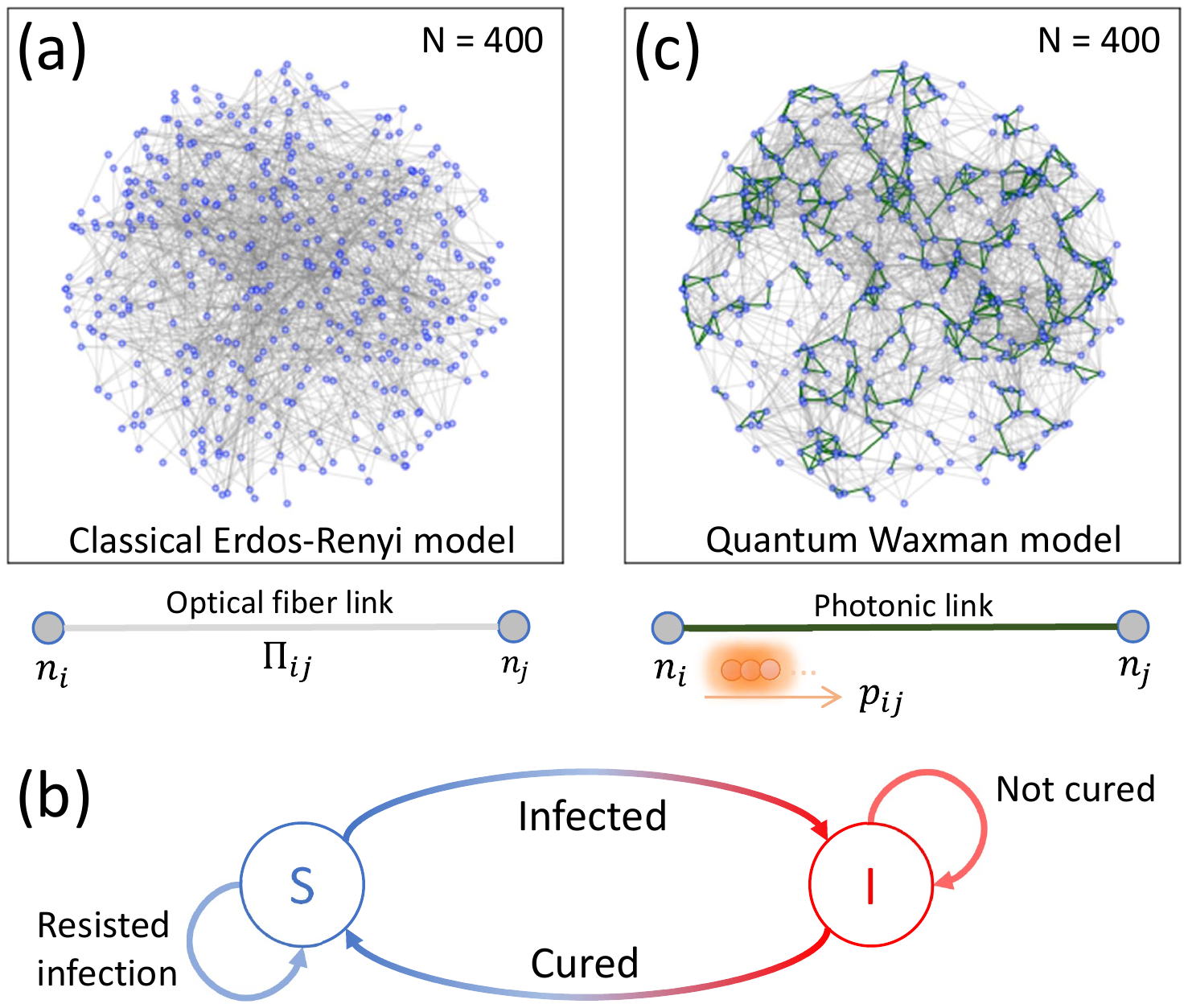}
	\caption{\label{FIG:fig0}
		\textbf{Network topologies and the SIS epidemiological model.}
		\textbf{(a)} A classical network generated using the Erdős–Rényi graph model, where the probability that two nodes $n_i$ and $n_j$ are connected by an optical fiber is $\Pi_{ij}=p=0.01$.
		\textbf{(b)} Schematic illustration of the susceptible–infected–susceptible (SIS) model describing the infection dynamics of a single node. The two possible states, S (susceptible) and I (infected), represent recovery and reinfection processes within the network.
        \textbf{(c)} A quantum network generated using the Waxman model. Dark green lines denote optical-fiber photonic links, established by transmitting multiple single photons (depicted as glowing orange disks) between nodes, with an overall success probability $p_{ij}$.
    }
\end{figure}

{\color{blue}\emph{Quantum network built upon photonic link.}} Before adopting an epidemiological model, it is essential to first establish the key physical and statistical characteristics of the quantum network. For simplicity, we restrict our analysis to undirected graphs comprising $N$ nodes (representing end users), where each pair of nodes $n_i$ and $n_j$ is connected with probability $\Pi_{ij}$. A simple example is the Erdős–Rényi (ER) graph model, shown in Fig.~\ref{FIG:fig0}(a), whereh $\Pi_{ij}=p$ is a constant. 
A quantum network is fundamentally determined by two components: (1) the classical network topology (i.e., $\Pi_{ij}$) and (2) the mode of quantum communication. 

We consider a quantum network built upon the existing classical optical fiber infrastructure, whose topology is modeled using the Waxman random graph model. In this model, the probability of forming a link between two nodes $n_i$ and $n_j$ separated by a distance $d_{ij}$ is given by 
 $\Pi_{ij}=\beta_L  e^{(-d_{ij}/\alpha_L)}$, where the parameters $\alpha_L>0$ and $0<\beta_L\leq1$ control the average link length and the mean node degree, respectively. Here we adopt representative parameters for the U.S. fiber-optic network, namely $\alpha_L=226$~km and $\beta_L=1$ \cite{Lakhinageographic2003, DurairajanIntertubes2015}, with all nodes uniformly distributed within a circular area of radius $R_\text{max}=1600$~km.
 
The quantum feature is incorporated through a photonic quantum network model following Ref.~\cite{BritoStatistical2020}, in which nodes connected by an optical fiber can establish quantum communication via probabilistic single photon transmission. An example of such a network is illustrated in Fig.~\ref{FIG:fig0}(c).
Once $n_i$ and $n_j$ are connected by an optical fiber link (with a probability $\Pi_{ij}$), the probability of successfully transmitting a single photon (e.g., one photon from an entangled pair) is given by $P_{ij}=10^{-\gamma d_{ij}/10}$, where $\gamma$ denotes the fiber attenuation coefficient, which depends on the photon wavelength. If
$n_p$ photons are sent in each attempt to establish communication, the probability of successfully receiving at least one photon, i.e., achieving a successful quantum link between the two nodes, is
\begin{equation}
    p_{ij} = 1 - (1-P_{ij})^{n_p}.
\end{equation}
We refer to this network framework as the quantum Waxman graph model. The physical characteristics of the optical fiber therefore determine the statistical behavior of the quantum network, resulting in richer and more complex spreading dynamics, as discussed later. Following state-of-the-art experimental parameters, we use
$n_p=1000$ and $\gamma=0.2$dB/km, which are typical values for optical fibers operating at a wavelength of 1550 nm \cite{BritoStatistical2020}.

{\color{blue}\emph{Virus spreads in classical networks.}} We review a class of epidemiology-inspired models commonly used to study virus propagation in classical networks, originating from the seminal Kephart–White (KW) model. This framework employs the standard SIS (susceptible $\to$ infected $\to$ susceptible) epidemiological process, where an individual becomes susceptible again immediately after recovery, as illustrated in Fig.~\ref{FIG:fig0}(b).

The Kephart–White (KW) model assumes that when an edge exists from node $n_i$ to another node $n_j$, the latter can be infected by the former. Each edge and node is associated with an infection rate $\beta$ and a curing rate $\delta$. For sufficiently large networks, the fraction of infected nodes at time $t$, denoted by $\eta(t)$, can be treated as a continuous variable. Under the assumption that $\beta$ and $\delta$ are constants, its dynamics are governed by
\begin{equation}
\dot{\eta}(t)=\beta\langle k\rangle\eta(t)(1-\eta(t))-\delta \eta(t),
\end{equation}
where $\langle k\rangle$ represents the average node degree or connectivity. For example, for an ER graph, this is given by $\langle k\rangle=p(N-1)$. The solution to the above equation is
$\eta(t) = \frac{\eta_{0}\eta_{\infty}}{\eta_{0}(\eta_{\infty}-\eta_{0})e^{-(\beta \langle k\rangle-\delta)t}}$, where $\eta_{0}=\eta(0)$ and $\eta_{\infty}=\lim_{t\to\infty}\eta(t)$ are constants. 
Importantly, the model predicts a critical point known as the \textit{steady-state epidemic threshold} $\tau_c$, which determines whether an infection can persist in the network. A viral outbreak quickly dies out if $\tau=\beta/\delta < \tau_c$, making this parameter essential for predicting the long-term stability of the system. For the KW model, the threshold is given by $\tau_{\text{KW}} = 1/\langle k\rangle$.

Although the KW model provides useful approximations, real-world networks are typically inhomogeneous, and the node degree $k$ may follow a complex distribution $p_k$. Several efforts have been made to establish more general theory of virus spreading in classical networks. One such approach is the mean-field approximation (MFA), which treats all graphs sharing the same $p_k$ as statistically equivalent and predicts an epidemic threshold of $\tau_\text{MFA} = \frac{\langle k\rangle}{\langle k^2\rangle}$ \cite{SatorrasEpidemic2002}. 

A more sophisticated framework is the nonlinear dynamical system (NLDS) model, in which the epidemic threshold cannot be derived solely from the degree distribution $p_k$. Instead, it depends explicitly on the detailed network topology, typically characterized by the adjacency matrix (AM) $A$. In this framework, the epidemic threshold is given by $\tau_\text{AM} = 1/\lambda_{1,\text{A}}$, where $\lambda_{1,\text{A}}$ denotes the largest eigenvalue of $A$. The NLDS model has been successfully applied to a variety of classical networks, including the Oregon network, the Barabási–Albert scale-free model, and the Erdős–Rényi (ER) model \cite{WangEpidemic2003}.

The NLDS approach describes infection dynamics as a Markov process, where the network configuration at time step $t$ depends only on that at the previous time step $t-1$, assuming sufficiently small temporal intervals. However,since the full Markov chain grows exponentially with the number of nodes, direct computation becomes intractable. To overcome this, the NLDS model employs an independence assumption, treating the infection probabilities $p_{i,t}$, the probability that node $n_i$ is infected at time $t$, as statistically independent. Under this assumption, the evolution of the infection probability can be written as
\begin{equation} \label{equ_main}
    p_{i,t} = 1-(1-p_{i,t-1})\xi_{i,t} - \delta p_{i,t-1}\xi_{i,t},
\end{equation}
where $\xi_{i,t} = \prod_{\langle j,i\rangle} (1 - \beta p_{j,t-1})$ is the probability that node $n_i$ receive no infections from any of its neighbors at the next time-step, with $\langle j,i\rangle$ taken over all node $n_j$  connected to $n_i$.

\begin{figure}[h]
    \includegraphics[width=1.0\columnwidth]{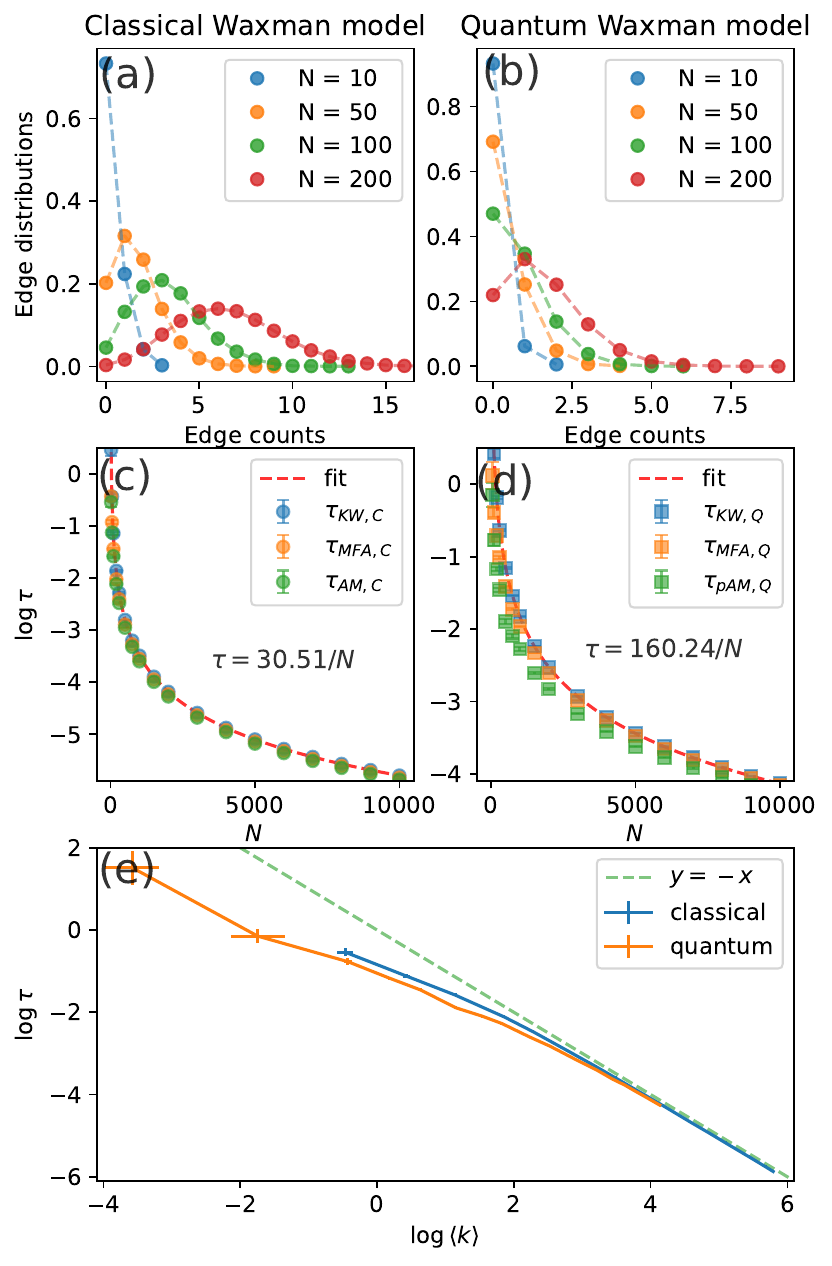}
	\caption{\label{FIG:fig1}
		\textbf{Edge distributions and epidemic thresholds of the classical and quantum networks.}
		\textbf{(a)} and \textbf{(b)}~Degree distribution $p(k)$ with varying numbers of nodes $N$ in classical and quantum networks based on the Waxman topology.
            \textbf{(c)} and \textbf{(d)}~Epidemic thresholds $\tau_\text{KW}$, $\tau_\text{MFA}$ and $\tau_\text{AM}$ ($\tau_\text{pAM}$ for quantum case) versus $N$ , computed numerically using Monte Carlo simulations with 200 random realizations for each $N$. For large $N$, all thresholds converge and exhibit the scaling $\tau \approx c/N$, with fitted constants $c=30.51\pm0.03$ and $c=160.24\pm0.11$ for classical and quantum networks.
            \textbf{(e)}~Epidemic threshold versus average connectivity in log-log scale. Both classical and quantum models approach the asymptotic relation $\log\tau = -\log\langle k\rangle$.
 }
 \end{figure}

{\color{blue}\emph{Epidemic thresholds of quantum networks}}. 
The quantum Waxman graph model enables us to investigate the resilience of the quantum network under viral spreading dynamics and to compare its behavior with that of the classical network. We first examine the edge-length distributions $p_k$ of quantum and classical networks, as shown in Figs.~\ref{FIG:fig1}(a) and (b). As the network density increases, the distribution peak shifts toward larger values, indicating that each node becomes more likely to connect with nearby neighbors. Overall, owing to finite photon loss in optical fibers, the connectivity of a quantum network is sparser than that of its classical counterpart with the same underlying topology.

In the classical Waxman graph, both thresholds $\tau_\text{KW}$ and $\tau_\text{MFA}$ can be computed from the degree distribution $p_k$ while the exact value of $\tau_{AM}$ needs to be evaluated numerically via Monte Carlo simulation  of the NLDS model (see Supplemental Material). The three thresholds for the classical Waxman graph are plotted in Fig.~\ref{FIG:fig1}(c), showing good agreement and convergence to the same value for large $N$. Importantly, the thresholds exhibit a scaling behavior $\tau \propto \frac{1}{N}$, which follows naturally from the scaling $\langle k\rangle \propto N$ in the classical backbone network.  

Similarly, the quantum thresholds $\tau_\text{KW}$ and $\tau_\text{MFA}$ can be computed from $p_k$ for the corresponding quantum Waxman graph. In classical networks, the adjacency matrix $A$ contains only binary entries (0 and 1), but this is no longer true for the quantum network, where the success rate of the photonic link $p_{ij}$ introduces weighted connections. To account for this, we define a probability adjacency matrix (pAM) by assigning each edge a weight $p_{ij}$ if nodes $n_i$ and $n_j$ are connected and 0 otherwise. The epidemic threshold for the quantum network, denoted $\tau_\text{pAM}$, can then be computed within the NLDS framework, since virus spreading across $h$ hops can still be approximated as $(\beta A)^h$.

Fig.~\ref{FIG:fig1}(d) presents the results for the quantum Waxman model. We first note that the scaling $\tau_\text{KW} \propto \frac{1}{N}$ remains valid, albeit with a larger prefactor, as it reflects only the average connectivity. Both $\tau_\text{MFA}$ and $\tau_{pAM}$ deviate from $\tau_\text{KW}$ at small $N$ but converge as $N$ becomes sufficiently large. Notably, $\tau_{pAM}$ does not generally scale as $\frac{1}{N}$, indicating the emergence of nontrivial corrections in the quantum network due to probabilistic photon links. These effects, however, vanish in the thermodynamic limit $N \to \infty$.

Direct comparison between the thresholds further reveals that quantum networks typically exhibit higher epidemic thresholds due to their effectively sparser connectivity. To enable an apples-to-apples comparison, we plot both the average connectivity and the thresholds ($\tau_{AM}$ and $\tau_{pAM}$) on a log–log scale in Fig.~\ref{FIG:fig1}(d). At large $\langle k\rangle$ (corresponding to large $N$), both the classical and quantum networks follow the scaling relation $\log\tau = -\log\langle k\rangle$, demonstrating that $\tau \langle k\rangle \to 1$ in the thermodynamic limit also holds for the quantum network, in the absence of localization or large degree variance.

\begin{figure}
	\includegraphics[width=0.97\columnwidth]{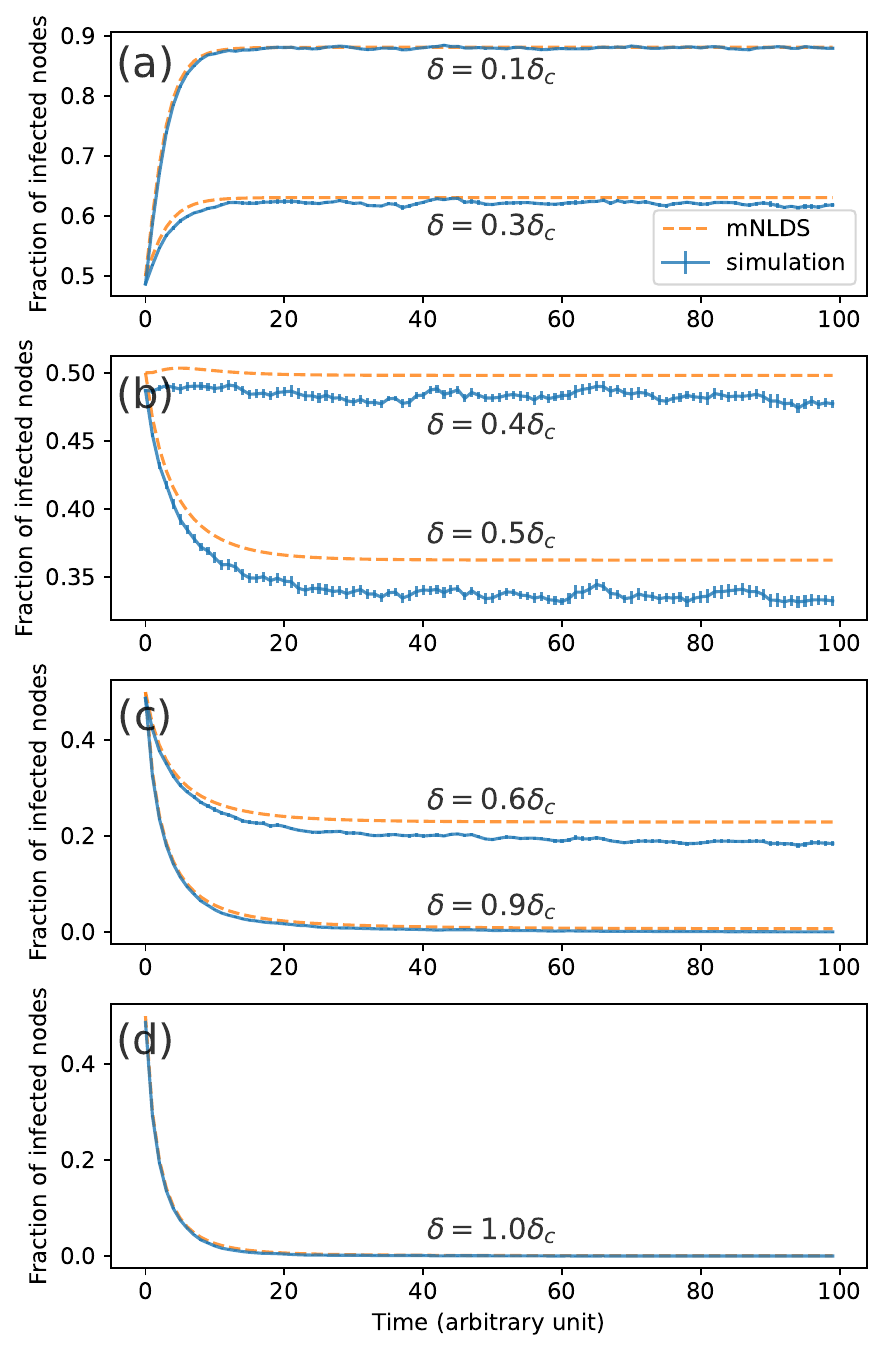}
	\caption{\label{FIG:fig2}
		\textbf{Epidemic dynamics in the quantum network.}
		\textbf{(a)} For small curing rares ($\delta=0.1\delta_c$ and $\delta=0.3\delta_c$, the quantum network remains partially infected with $\eta_\infty > \eta_0$.
        \textbf{(b)} Increasing the curing rate to $\delta=0.4\delta_c$ keeps $\eta(t)$ close to its initial value throughout the evolution, while at $\delta=0.5\delta_c$, $\eta_\infty < \eta_0$. In both cases, mNLDS and direct simulations show small discrepancies, and the latter exhibit larger run-to-run fluctuations.
        \textbf{(c)} For higher curing rates ($0.6\delta_c$ and $\delta=0.9\delta_c$), $\eta(t)$ rapidly decreases, and the two methods yield more consistent results.
        \textbf{(d)} At the critical point $\delta=\delta_c$, both approaches predict the infection vanishes asymptotically ($\eta_{\inf} \to 0$).
       Simulations are performed on a quantum network with 2000 nodes and an initial infection probability $p_{i,0}=0.5$. The infection rate is $\beta=0.05$, and direct simulations are averaged over 20 independent runs with random initial configurations.
    }
\end{figure}

{\color{blue}\emph{Dynamics of virus spreading in quantum networks.}} 
The quantum threshold $\tau_{pAM}$ represents only one aspect of the resilience of a quantum network. To further capture the temporal dynamics of viral spreading, we develop a modified nonlinear dynamical system (mNLDS) model tailored to the quantum network. To maintain consistency with the framework introduced in Eq.~(\ref{equ_main}), we incorporate the probability adjacency matrix $A$ to describe the infection dynamics. Accordingly, we define
\begin{equation} \label{equ_adj}
    \xi_{i,t} = \prod_j (1 - \beta A_{ij} p_{j,t-1}),
\end{equation}
where $\xi_{i,t}$ naturally reduces to the classical form when the entries of $A$
are binary. We refer to this extended formalism as the modified NLDS (mNLDS) model, and we use it throughout the following analysis.

We emphasize that the mNLDS model is applicable to the quantum networks considered in this work, as the independence assumption continues to hold. However, in more advanced quantum network architectures, where nodes function as quantum processing units or include quantum memories capable of storing entangled states, long-range entanglement can develop between distant nodes. In such scenarios, the independence assumption no longer applies, and a full simulation of the underlying Markovian processes becomes necessary. Accurately modeling these correlated dynamics would therefore require more sophisticated algorithms and substantial computational resources.

As a demonstration, we employ the mNLDS framework to simulate the infection dynamics of a quantum network and verify that it reliably captures the spreading behavior. The corresponding threshold $\tau_{pAM,Q}$ also provides an accurate prediction of the epidemic threshold for the quantum network. Specifically, we generate a random instance of the Waxman model containing 2000 nodes and perform both mNLDS simulations and direct stochastic simulations of the infection process. The infection rate is fixed at $\beta = 0.05$, and the critical curing rate is defined as $\delta_c = \beta/\tau_{pAM,Q}$. To visualize the transition across the threshold, we vary the curing rate $\delta$ and plot the resulting temporal dynamics in Fig.~\ref{FIG:fig2}. The initial infection probability is set to $\eta_0=p_{i,0}=0.5$. Similar to classical networks, the steady-state infection density $\eta_{\infty}$ converges and is largely insensitive to the specific configuration of the initial state, although transient dynamics may differ (see Supplemental Materials).

When the curing rate is smaller than the critical value, we expect $\eta_{\infty}$ to converge to a non-zero value. As shown in Fig.~\ref{FIG:fig2}(a), for $\delta\leq0.3\delta_c$, the results from mNLDS and direct simulations coincide almost perfectly, and the network evolves toward a state with a higher fraction of infected nodes than the initial condition. Increasing the curing rate to
$\delta$ to $0.4\delta_c$, as illustrated in Fig.~\ref{FIG:fig2}(b), leads to a noticeable deviation between the two approaches. In this regime, the direct simulations converge more slowly and exhibit larger fluctuations. Interestingly, $\eta(t)$ remains close to its initial value, suggesting that infection and recovery processes are nearly balanced. Further increasing the curing rate to $\delta$ to $0.5\delta_c$ and $0.6\delta_c$ [see Figs.~\ref{FIG:fig2}(b) and (c)] results in $\eta_\infty<\eta_0$, with the steady-state infection density continuing to decrease.
As $\delta$ approaches $\delta_c$, both methods consistently show that $\eta_\infty \to 0$. At the critical point $\delta=\delta_c$, mNLDS and direct simulations match precisely and converge to zero [Fig.~\ref{FIG:fig2}(d)]. These results demonstrate that the mNLDS model accurately captures the dynamical behavior and epidemic threshold of the quantum network under viral infection.

{\color{blue}\emph{Discussions and conclusions.}}
While we have explored several intriguing aspects of virus spreading in quantum networks, many open questions remain. In this work, all nodes in the quantum network are treated as classical entities; however, the problem becomes far more complex when nodes themselves exhibit quantum behavior, for instance, when they include quantum memories or quantum processing units. In such cases, additional quantum effects, such as quantum viruses or long-range entanglement among nodes, could play a crucial role in shaping the spreading dynamics.

For simplicity, we have focused on a single mode of quantum communication, optical fiber links. In practice, multiple quantum communication channels already exist, including free-space and satellite-based quantum links for long-distance information exchange. Incorporating these hybrid communication methods would significantly enrich the structure of quantum networks and lead to qualitatively new infection dynamics and thresholds.

Another promising direction involves hybrid classical–quantum networks. As quantum communication technologies mature, classical communication will continue to coexist rather than be completely replaced. Realistic models must therefore account for mixed topologies \cite{Feng2025}, heterogeneous node behaviors, and potentially node-dependent infection and curing rates. More advanced classical frameworks, such as the N-intertwined Markov chain model \cite{MieghemVirus2009}, may serve as valuable starting points for such hybrid analyses.

In conclusion, we have investigated the fundamental question of how viral processes evolve within the quantum network. By combining the standard SIS epidemiological framework with a quantum Waxman graph model based on photonic fiber links, we examined the resilience of quantum networks under viral spreading. Using three distinct approaches, we revealed intrinsic structural differences between quantum and classical networks through comparative analyses of their epidemic thresholds. Furthermore, we proposed a modified nonlinear dynamical system (mNLDS) model capable of capturing the temporal dynamics of viral spreading across probabilistic quantum links.

Our results provide a foundational step toward understanding epidemic processes in quantum communication infrastructures and offer valuable insights for the secure and resilient design of large-scale quantum networks. This work bridges the emerging intersection of quantum information science, network theory, and epidemiology, paving the way for a future quantum epidemiological theory.

\begin{acknowledgments}
{\color{blue}\emph{Acknowledgements.}} J. Hou thanks C. Lu for inspiring discussions. This work is supported by the Air Force Office of Scientific Research under Grant No. FA9550-20-10220 and the National Science Foundation under Grant No. PHY-2409943, OSI-2228725. C.Z. is supported by
Department of Energy (DE-SC0022069).
\end{acknowledgments}

\newpage \clearpage
\onecolumngrid
\appendix

\setcounter{table}{0} \renewcommand{\thetable}{A\arabic{table}} %
\setcounter{figure}{0} \renewcommand{\thefigure}{A\arabic{figure}}

\section{Supplemental materials for ``Virus Spreading in Quantum Networks"}
In this Supplemental Material, we provide additional details on the computation of epidemic thresholds, analyze how physical properties influence these thresholds, and present the derivation of the equation of motion governing virus spreading in a quantum network.

\subsection{Computing the thresholds for classical and quantum networks} \label{compute_edge_dis_and_thresholds}
In the main text, we analyze several epidemic thresholds for both classical and quantum network models using analytical and numerical methods. Here, we provide additional methodological details and discuss the physical interpretations underlying the different computational schemes.

\subsubsection{Classical models}
For the classical Kephart–White (KW) model and the mean-field approximation (MFA), the epidemic thresholds can be expressed analytically as
\begin{equation}
\tau_{\text{KW}} = \frac{1}{\langle k \rangle}, \qquad 
\tau_{\text{MFA}} = \frac{\langle k \rangle}{\langle k^2 \rangle},
\end{equation}
where $\langle k \rangle = \sum_k k\,p_k$ and $\langle k^2 \rangle = \sum_k k^2\,p_k$ are the first and second moments of the degree distribution $p_k$ \cite{KephartDirected1991,SatorrasEpidemic2002}.  
Here $p_k$ is normalized as $\sum_k p_k = 1$.  
In the special case of a regular graph, where the degree variance vanishes ($\langle k^2\rangle = \langle k\rangle^2$), the two thresholds coincide, i.e., 
$\tau_{\text{KW}}\equiv\tau_{\text{MFA}}$.

Unlike these degree-based (annealed) thresholds, the adjacency-matrix (AM) or spectral threshold accounts for the complete topology of a specific network realization.
Let $A$ be the adjacency matrix, with entries $A_{ij}=1$ if nodes $i$ and $j$ are connected and $A_{ij}=0$ otherwise.  
Denoting the largest eigenvalue of $A$ by $\lambda_{1,A}$, the spectral threshold is given by
\begin{equation}
\tau_{\text{AM}} = \frac{1}{\lambda_{1,A}},
\end{equation}
as established in Ref.~\cite{WangEpidemic2003}.  
This threshold follows from linearizing the infection dynamics around the disease-free state, where the growth of infection probability over $h$ propagation steps is governed by $(\beta A)^h$ and dominated by the spectral radius $\lambda_{1,A}$.  
Consequently, $\tau_{\text{AM}}$ captures quenched topological effects that are inaccessible through degree-distribution-based approximations.

\subsubsection{Quantum network models}
In quantum networks, node connectivity depends not only on the underlying optical-fiber topology but also on the probabilistic nature of photon transmission.
We denote by $\Pi_{ij}$ the probability that a fiber physically connects nodes $i$ and $j$, and by $p_{ij}$ the probability for successfully establishing a quantum link across that fiber.  
Thus, the effective connection between two nodes is determined jointly by $\Pi_{ij}$ and $p_{ij}$.  
Different ways of combining these two probabilities yield distinct definitions of epidemic thresholds, each corresponding to a different physical regime of disorder. 
We consider three representative methods:

\paragraph*{Method 1 (Annealed approximation).}
The simplest approach replaces stochastic connectivity by its ensemble average, constructing an ``expected adjacency'' matrix 
\begin{equation}
M^{(1)}_{ij} = \Pi_{ij} \, p_{ij}.
\end{equation}
The corresponding epidemic threshold is then computed as $\tau_{\text{AM}} = 1/\lambda_1(M^{(1)})$.  
This approach represents an \emph{annealed} regime, in which both the fiber layout and the photonic links fluctuate rapidly on timescales comparable to or shorter than the epidemic dynamics.  
In this limit, the spreading process effectively samples the average, weighted network.

\paragraph*{Method 2 (Quenched topology, annealed photonics).}
In this approach, we first generate a binary adjacency matrix $A_{ij}\!\in\!\{0,1\}$ by random sampling according to $\Pi_{ij}$, representing a fixed realization of the fiber infrastructure.  
We then assign link weights $p_{ij}$ to each existing edge and construct
\begin{equation}
M^{(2)}_{ij} = A_{ij} \, p_{ij}.
\end{equation}
This scheme models a \emph{quenched} topology (static fiber connections) with stochastic but stationary photonic transmissivities.  
The corresponding spectral threshold $\tau_{\text{AM}} = 1/\lambda_1(M^{(2)})$, captures the influence of topological variability across different network realizations.

\paragraph*{Method 3 (Fully sampled, quenched links).}
In this approach, both the fiber layout and the photonic connection are explicitly sampled.  
For each network instance $A_{ij}$ generated from $\Pi_{ij}$, we perform an independent Bernoulli sampling of each link with success probability $p_{ij}$, producing a binary matrix $B_{ij}$.  
The effective adjacency used for threshold computation is the elementwise product
\begin{equation}
M^{(3)}_{ij} = A_{ij} \, B_{ij}.
\end{equation}
This method represents a \emph{fully quenched} system, appropriate when both the physical fiber layout and the photonic links are random on the timescale of interest (e.g., per transmission attempt).

\begin{figure}[h]
\includegraphics[width=0.48\columnwidth]{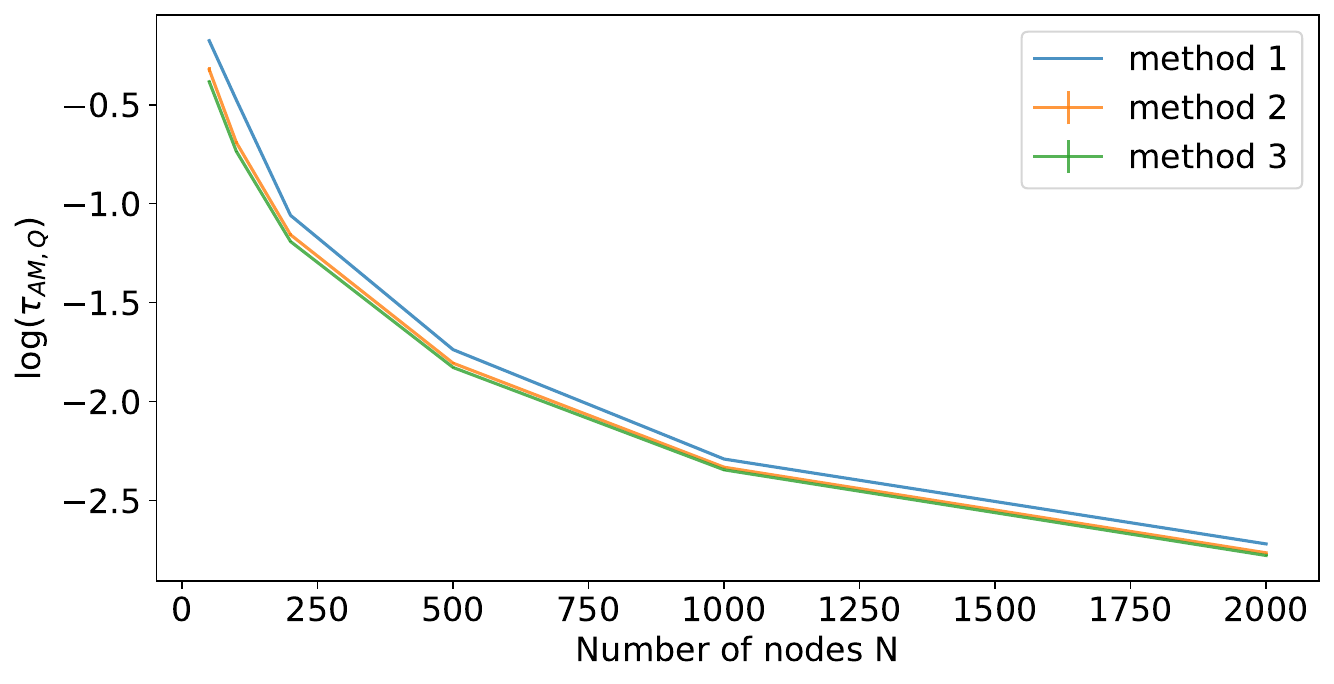}
	\caption{\label{FIG:threshold_methods}
		\textbf{Thresholds from different methods.} We apply the three methods discussed to compute the threshold $\tau_\text{AM}$ for quantum networks with varying numbers of nodes $N$. In Method 2, we sampled 200 instances of the Waxman graph for each size $N$ to compute the mean and variance of the threshold holds of each instance. In Method 3, we further sample 100 instances for each instance in Method 2. All parameters are the same as those used in the main text.
    }
\end{figure}

\subsubsection{Comparison and interpretation}
Because the spectral radius is a nonlinear function of the matrix entries, in general
\begin{equation}
\lambda_1\!\big(E[M]\big) \neq E\!\big[\lambda_1(M)\big],
\end{equation}
and therefore the thresholds obtained from Methods 1--3 need not coincide.  
Empirically we find that Methods 2 and 3 yield closely consistent results for the parameter ranges considered in this work, whereas Method 1 exhibits systematic deviations, as shown in Fig.~\ref{FIG:threshold_methods}. 
This discrepancy arises because Method 1 uses the eigenvalue of the mean matrix, thereby neglecting the correlations and structural heterogeneities present in individual graph realizations.  
In contrast, Methods 2 and 3 explicitly incorporate quenched topological disorder and more accurately represent the physical scenario of a quantum network with a fixed underlying fiber topology.  

The small difference between Methods 2 and 3 indicates that, within the parameter range considered, replacing binary photonic realizations with their mean transmissivities ($p_{ij}$) provides an accurate approximation.  
However, for sparser networks or small $p_{ij}$ values, Method 3 can exhibit larger fluctuations in $\tau_{\text{AM}}$, as rare successful links dominate the spectral properties.  
We quantify this variability by averaging the computed thresholds over many random instances and reporting both the mean and the standard deviation.  

Throughout this work, we employ Method 2 to compute the nonlinear dynamical system (NLDS) epidemic threshold, as it captures the static nature of the fiber topology while incorporating probabilistic photonic connectivity.  
Method 3 is used when computing edge-level statistics such as degree distributions or connectivity probabilities, unless stated otherwise.

\subsubsection{Physical interpretation and scaling}
The three methods described above correspond to distinct physical regimes of temporal and structural disorder:  
(i) Method 1—annealed (fast-fluctuating) disorder;  
(ii) Method 2—quenched topological disorder with averaged photonic weights; and  
(iii) Method 3—fully quenched disorder.  
In dense or homogeneous regimes, these approaches converge as $N$ increases and the spectral radius concentrates.  
In sparse or heterogeneous networks, however, $\lambda_1$ may localize on high-degree nodes or tightly connected clusters, leading to significant discrepancies among the methods.  
These differences underscore the importance of specifying the relevant timescale hierarchy between network fluctuations and epidemic dynamics when defining thresholds for quantum-network spreading processes.

\begin{figure}[h]
\includegraphics[width=0.75\columnwidth]{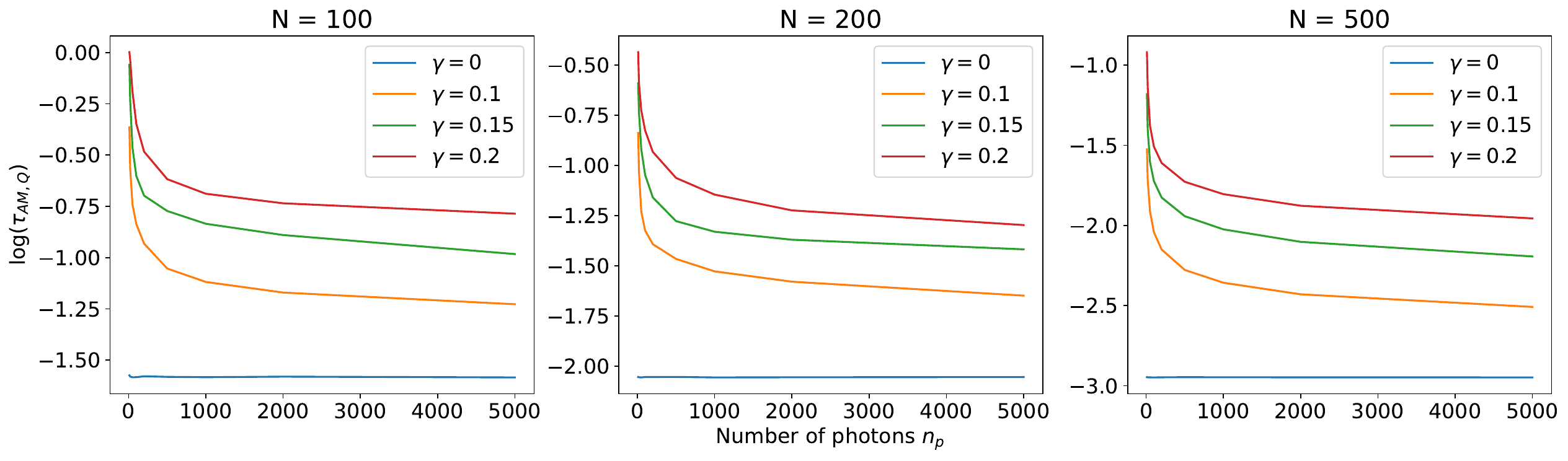}
	\caption{\label{FIG:threshold_pl_effects}
		\textbf{Thresholds versus $n_p$.} In each panel, the epidemic thresholds of quantum networks are computed for varying photon loss cofficients and the number of photons used in a single communication. From left to right, the panels correspond to increasing numbers of nodes in the quantum network. In all the simulations, we fix $R_\text{max} = 1600$ km, $\alpha_L = 216$ km and $\beta_L=1$. Each data point represents an average over 1000 random instances.
    }
\end{figure}

\subsection{Effects of physical properties on epidemic thresholds}
In the main text, our discussion of the quantum network was restricted to fixed parameter values, including the photon loss coefficient $\gamma=0.2$, the number of photons used per communication $n_p=1000$ and the network radius $R_\text{max}=1600$km. In this section, we examine how these parameters—and, more broadly, the physical properties of the quantum network—affect the epidemic thresholds.
We further show that the threshold becomes asymptotically independent of these parameters as the network size increases.

For the photon loss, the intrinsic attenuation limit of the silica optical fibers is estimated to lie between 0.095 to 0.13 dB/km \cite{TsujikawaIntrinsic2005}. Accordingly we choose $\gamma = 0.1$, 0.15 and 0.2 for the following simulations. We also use $\tau_\text{AM}$ as the epidemic threshold.

\begin{figure}[h]
\includegraphics[width=0.75\columnwidth]{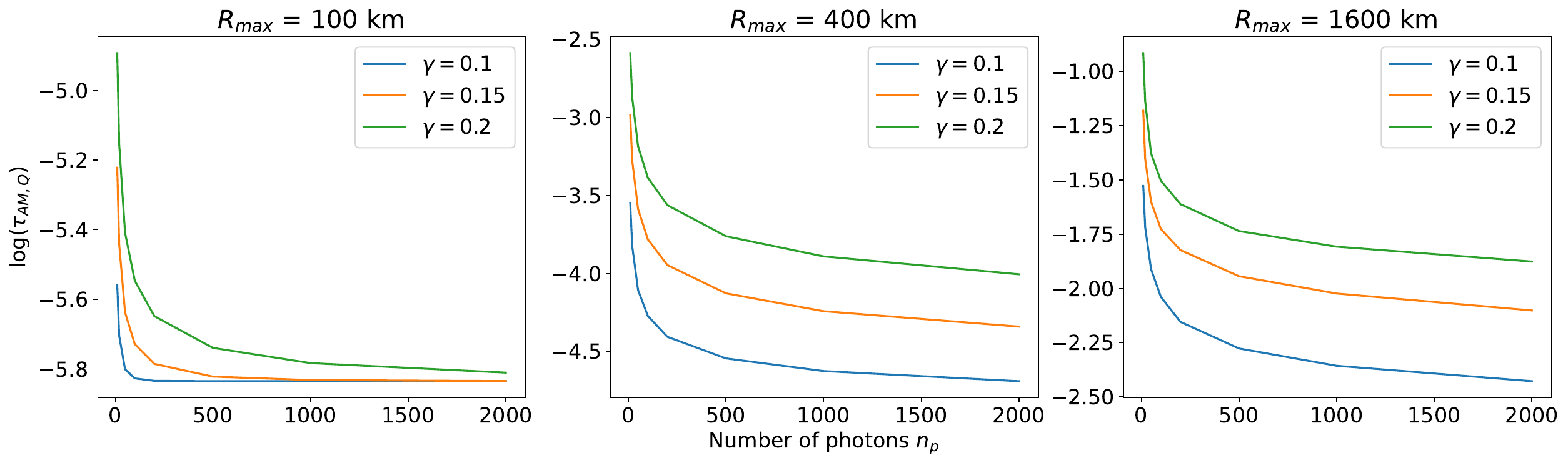}
	\caption{\label{FIG:threshold_pl_radii}
		\textbf{Thresholds versus $n_p$.} In each panel, the epidemic thresholds of quantum networks are computed for varying photon loss coefficients and number of photons used per communication attempt. From left to right, the panels correspond to increasing numbers of nodes in the quantum network. In all the simulations, we fix $N = 500$, $\alpha_L = 216$ km and and $\beta_L=1$. Each data point represents an average over 1000 random instances. 
    }
\end{figure}

In Fig.~\ref{FIG:threshold_pl_effects}, we first examine how the epidemic thresholds depend on $\beta_L$ and $n_p$ for different network sizes $N$. In all cases, the thresholds decrease as more photons are transmitted per communication attempt.
This behavior is expected, since the “virus,” analogous to information, propagates more readily when the success probability of a single communication increases. When $n_p$ just starts to increase, the thresholds drop rapidly even on a log scale, but beyond a certain turning point, the decrease becomes approximately linear. This trend persists across different network sizes $N$.

When we compare the curves for different photon loss coefficients, we do not observe any significant differences as $N$ changes from 100 to 500. This indicates that fiber loss has only a minor effect on the epidemic thresholds in a quantum network. However, certain secondary effects of reducing fiber loss may not be captured in these plots. For instance, lowering fiber attenuation could enable connections between more distant nodes, thereby modifying the distance distribution $\Pi_{ij}$.

Fig.~\ref{FIG:threshold_pl_radii} is similar to Fig.~\ref{FIG:threshold_pl_effects}, except that we fix $N=500$ and vary $R_\text{max}$, which represents the geographical span of the quantum network. Again, photon loss barely affects the threshold. When the network is small, or $R_\text{max} < \alpha_L$, the threshold decreases rapidly and approaches a saturated value as $n_p$ increases. For larger networks with $R_\text{max} > \alpha_L$, the curves become less steep.

\subsection{Simulating the dynamics of epidemic spreading in quantum networks}
\subsubsection{Derivation of the equations of motion}
In the main text, we introduced a modified nonlinear dynamical system (mNLDS) model, Eqs.~(\ref{equ_main}) and~(\ref{equ_adj}), to describe the dynamics of epidemic spreading in quantum networks.  
Here we provide a detailed derivation of these equations and clarify the underlying assumptions.

We consider a quantum network composed of classical computing devices that communicate through quantum channels.  
Because the end nodes are classical, the global system state can be represented by a bit string $s_I$ of length $N$, where $s_I(i)=1$ indicates that node $n_i$ is infected and $s_I(i)=0$ otherwise.  
The configuration $s_0 = 00\cdots0$ denotes a healthy network, while $s_{2^N-1} = 11\cdots1$ represents a full infected one.  
At any discrete time step $t$, if the network is in state $s_I$, the probability of transitioning to another configuration $s_J$ at time $t+1$ is given by the conditional probability $P(s_J|s_I)$, which depends only on the current state and not on the prior history.  
This property defines a Markov chain that, in principle, fully determines the system’s dynamics.  
However, since the state space grows exponentially ($2^N$),  explicit computation rapidly becomes infeasible for large $N$.

To overcome this limitation, we adopt the standard \emph{independence assumption}, under which the infection probabilities of individual nodes are treated as statistically independent.  
Let $p_{i,t}$ denote the probability that node $n_i$ is infected at time $t$.  
This assumption is well justified for networks without quantum entanglement between nodes—i.e., the case relevant to the quantum communication model discussed in the main text.  
In the large-$N$ limit, the expected fraction of infected nodes is approximated as
\begin{equation}
\rho_t \simeq \frac{1}{N}\sum_i p_{i,t}.
\end{equation}
Hence, the epidemic dynamics reduce to determining the evolution of $\{p_{i,t}\}$ from one time step to the next.

Let $\xi_{i,t}$ denote the probability that node $n_i$ \emph{was not infected} by any of its neighbors at time $t$.  
For a neighbor $n_j$, the probability of transmitting the infection from $j$ to $i$ during one time step is $\beta A_{ij} p_{j,t-1}$, where $\beta$ is the infection rate and $A_{ij}$ is the (possibly weighted) adjacency-matrix element.  
Assuming independence among infection attempts from different neighbors, we obtain
\begin{equation}
\xi_{i,t} = \prod_j \bigl(1 - \beta A_{ij} p_{j,t-1}\bigr),
\label{eq:xi}
\end{equation}
which corresponds to Eq.~(\ref{equ_adj}) in the main text.

The probability that node $n_i$ is infected at time $t$ can then be expressed in terms of the complementary (non-infection) events.  
Two mutually exclusive cases lead to node $n_i$ being \emph{uninfected} at step $t$:
\begin{enumerate}
    \item The node was infected at $t\!-\!1$, did not receive new infections, and was cured.  
    The probability of this event is $\delta\, p_{i,t-1}\, \xi_{i,t}$, where $\delta$ is the recovery rate.
    \item The node was uninfected at $t\!-\!1$ and remained uninfected by all neighbors.  
    This occurs with probability $(1-p_{i,t-1})\, \xi_{i,t}$.
\end{enumerate}
Subtracting the sum of these probabilities from unity yields the update rule
\begin{equation}
p_{i,t} = 1 - \bigl[\delta p_{i,t-1} + (1-p_{i,t-1})\bigr]\xi_{i,t},
\label{eq:main_p}
\end{equation}
which reproduces Eq.~(\ref{equ_main}).  
Equation~(\ref{eq:main_p}) is a discrete-time nonlinear dynamical system that captures the evolution of node-level infection probabilities on a general weighted (classical or quantum) network described by $A$.  
For binary $A$, the model reduces to the standard epidemic dynamics on classical networks.  

Importantly, this derivation does not depend on the specific nature of the quantum communication channels.  
Any network characterized by a (possibly probabilistic) adjacency matrix $A$ can be described by the same formalism.

\subsubsection{Direct simulation using binary states}
To verify the validity of the probabilistic (mean-field) approximation above, we also perform direct stochastic simulations based on binary node states.  
Here each node $n_i$ is explicitly assigned a state $\sigma_i(t)\in\{0,1\}$ at each time step, representing healthy and infected conditions, respectively.  
While we still assume independence across nodes in generating events, this method captures the discrete realization of infections and recoveries rather than their probabilities.

At every time step $t$, the state $\sigma_i(t+1)$ is updated as follows:
\begin{enumerate}
    \item For each node, compute the probability of infection by its neighbors,
    \begin{equation}
    P_i^{(\text{inf})}(t) = 1 - \prod_j \bigl(1 - \beta A_{ij}\sigma_j(t)\bigr).
    \end{equation}
    A random number $r_i^{(\text{inf})}\in[0,1]$ is drawn; if $r_i^{(\text{inf})} < P_i^{(\text{inf})}(t)$, node $i$ is marked as infected by its neighbors.
    \item Independently, draw another random number $r_i^{(\text{cure})}$ to determine recovery.  
    If $\sigma_i(t)=1$ and $r_i^{(\text{cure})}<\delta$, the node is marked as cured.
    \item The final state is then updated according to these outcomes:  
    an infected node becomes healthy if it is cured and not reinfected, while a healthy node becomes infected if it was successfully infected by any neighbor.
\end{enumerate}
This algorithm yields a Monte-Carlo realization of the same underlying stochastic process represented by Eq.~(\ref{eq:main_p}).  
By averaging over many realizations, one recovers the expected trajectories of $\{p_{i,t}\}$ from the mNLDS model, confirming the consistency between the mean-field approximation and explicit binary-state dynamics.


\end{document}